\documentclass[a4paper,11pt]{article}
\pdfoutput=1 

\usepackage{jheppub} 

\usepackage[T1]{fontenc} 

\usepackage{subfigure}
\usepackage{graphicx,amssymb,amsmath,amsthm,amsfonts,epsfig,epsf}
\usepackage[usenames]{color}
\definecolor{darkred}{rgb}{0.5,0,0}

\usepackage{array}
\usepackage{afterpage}
\usepackage{bm}
\usepackage{dcolumn}
\usepackage[latin1,utf8]{inputenc}
\usepackage{latexsym}
\usepackage{rotating}
\usepackage{longtable}

\usepackage{enumerate}
\usepackage{tensor,multirow}
\usepackage{url}
\usepackage{placeins}

\usepackage{float}
\usepackage{graphicx}

\newcounter{mnotecount}[section]

\renewcommand{\themnotecount}{\thesection.\arabic{mnotecount}}

\newcommand{\mnote}[1]
{\protect{\stepcounter{mnotecount}}$^{\mbox{\footnotesize
$
\bullet$\themnotecount}}$ \marginpar{
\raggedright\tiny\em
$\!\!\!\!\!\!\,\bullet$\themnotecount: #1} }

\title{\boldmath Strong Cosmic Censorship in higher-dimensional Reissner-Nordstr\"{o}m-de Sitter spacetime}

\author[1]{Hang Liu,}
\author[1]{Ziyu Tang,}
\author[2]{Kyriakos Destounis,}
\author[3,4]{Bin Wang,}
\author[5]{Eleftherios Papantonopoulos,}
\author[6,7]{and Hongbao Zhang}

\affiliation[1]{Collaborative Innovation Center of IFSA (CICIFSA), Shanghai Jiao Tong University, Shanghai 200240, China}
\affiliation[2]{CENTRA, Departamento de F\'{\i}sica, Instituto Superior T\'ecnico -- IST, Universidade de Lisboa -- UL,
Avenida Rovisco Pais 1, 1049 Lisboa, Portugal}
\affiliation[3]{Centre for Gravitation and Cosmology, Yangzhou University, Yangzhou 225009, China}
\affiliation[4]{School of Aeronautics and Astronautics, Shanghai Jiao Tong University, Shanghai 200240, China}
\affiliation[5]{Physics Division, National Technical University of Athens, 15780 Zografou Campus, Athens, Greece}
\affiliation[6]{Department of Physics, Beijing Normal University, Beijing 100875, China}
\affiliation[7]{Theoretische Natuurkunde, Vrije Universiteit Brussel and The International Solvay Institutes, Pleinlaan 2, B-1050 Brussels, Belgium}

\emailAdd{hangliu@sjtu.edu.cn}
\emailAdd{tangziyu@sjtu.edu.cn}
\emailAdd{kyriakosdestounis@tecnico.ulisboa.pt}
\emailAdd{wang$\_$b@sjtu.edu.cn}
\emailAdd{lpapa@central.ntua.gr}
\emailAdd{hzhang@vub.ac.be}

\abstract{It was recently shown that Strong Cosmic Censorship might be violated for near-extremally-charged black holes in 4-dimensional de Sitter space under scalar perturbations. Here, we extend the study of neutral massless scalar perturbations in higher dimensions and discuss the dimensional influence on the validity of Strong Cosmic Censorship hypothesis. By giving an elaborate description of neutral massless scalar perturbations of Reissner-Nordstr\"{o}m-de Sitter black holes in $d=4,5$ and $6$ dimensions we conclude that Strong Cosmic Censorship is violated near extremality.}

\begin{document} 
\maketitle
\flushbottom
\section{Introduction}
It is well known that the would-be Cauchy horizon (CH) in asymptotically flat black holes (BHs) is a singular boundary \cite{Simpson:1973ua,Hartle,Poisson:1990eh}. The remnant fields of gravitational collapse have an inverse power-law decay behavior in the exterior
of asymptotically flat BHs \cite{Price1,Price2}, and will be amplified when propagating along the CH due to the
exponential blueshift effect occurring there. The gravitational collapse of matter fields cannot go beyond the CH to the timelike singularity in the eternal asymptotically flat BH, leading to the preservation of the deterministic power of physical laws and the Strong Cosmic Censorship (SCC) hypothesis, proposed by Penrose.

However, for de Sitter (dS) BH spacetimes, due to the change of the boundary conditions, remnant perturbation fields outside de Sitter BHs decay exponentially instead of polynomially \cite{Brady2, Brady3, Molina1, Abdalla1,Du:2004jt}.  The extendibility of the metric at the CH depends delicately on the competition
between the exponential decay outside the BH and the exponential blueshift
amplification along the CH. In such a scenario, the decay of perturbations could be fast enough such that the CH singularity can be so weak that the spacetime metric will be extendible beyond it as a weak solution to the Einstein field equations \cite{Chris1}, meaning that SCC may be violated!  Mathematically, it was proven \cite{Maeda1,Dafermos1,Costa1,Costa2,Costa3,Hintz1,Hintz2,Hintz3,Cardoso1}  that SCC will not be respected under massless neutral scalar perturbations if the following condition is satisfied
\begin{equation}
\beta\equiv-\frac{\mathrm{Im}(\omega)}{\kappa_{-}}>\frac{1}{2},\label{eq15}
\end{equation}
where $\kappa_{-}$ is the surface gravity of the CH and $\mathrm{Im}(\omega)$ is the imaginary part of the lowest-lying/dominant quasinormal mode (QNM) of the perturbation in the external region of the BH.

In particular, for near-extremal Reissner-Nordstrom-de Sitter (RNdS) BHs, it was shown \cite{Cardoso1} that neutral massless scalar perturbations are extendible past the CH, since the blueshift amplification along the CH is dwarfed by the exponential decay behavior outside of the dS BH. Such a violation of SCC
becomes even more severe in the case of the coupled electromagnetic and gravitational perturbations \cite{Dias3}. Later on, it was shown that the violation of SCC can be alleviated by considering a sufficiently charged scalar field on the exterior of RNdS BHs \cite{Hod1,Cardoso2,Zhang1,Dias1}, although there was still a region in the parameter space where SCC may be violated. Similar results have been obtained for Dirac field perturbations \cite{Zhang2,Destounis1}. On the other hand, the nonlinear evolution of massive neutral scalar fields in RNdS space revealed that SCC might not be saved by such nonlinear effects \cite{Luna1}. In addition, by investigating SCC in lukewarm RNdS and Martnez-Troncoso-Zanelli (MTZ) BH spacetimes, under non-minimally coupled massive scalar perturbations, it was demonstrated that the validity of the hypothesis depends on the characteristics of the scalar field \cite{Gwak1}.
Besides charged BHs, SCC has been examined in Kerr-de Sitter BH backgrounds and interestingly enough no violation was found for linear scalar and gravitational perturbations \cite{Dias2}.

All available studies of SCC in RNdS BH backgrounds are limited in 4 dimensions even though it has been found that in higher dimensions, physics becomes richer. In contrast to the 4-dimensional results,
various instabilities have been found in higher-dimensional spacetimes. In a wide class of $d\geq 4$ configurations, such as black
strings and black branes, the Gregory-Laflamme instability against linear perturbations was discussed in
\cite{Gregory1, Gregory2}. For higher-dimensional BHs in the Einstein-Gauss-Bonnet theory, it was found
that instabilities occur for large angular quantum numbers $l$, while the lowest lying QNMs of the first few angular quantum numbers were found stable \cite{Konoplya2,Beroiz1}. In particular,  numerical investigations have uncovered
the surprising result that RNdS BH backgrounds are unstable for $d\geq 7$, if the values of black hole mass and charge are large enough \cite{Konoplya1}, followed by the analytic confirmation of \cite{Cardoso3}. Moreover,  it was argued that the Weak Cosmic Censorship hypothesis could be restored easier in higher dimensions \cite{Mken} by examining the gravitational collapse of spherically symmetric generalized Vaidya spacetimes.
It is, thus, of great interest to generalize the discussion of SCC to higher-dimensional RNdS BHs and explore whether and how they affect the validity of the conjecture.

\section{Scalar fields in higher-dimensional RNdS spacetime}
The $d$-dimensional RNdS spacetime is described by the metric \cite{Chabab1}
\begin{equation}
\label{dspace}
ds^2=-f(r)dt^2+\frac{1}{f(r)} dr^2+r^2 d\Omega_{d-2}^2,
\end{equation}
where
\begin{align}\label{metric}
f(r)=1-\frac{m}{r^{d-3}}+\frac{q^2}{r^{2(d-3)}}-\frac{2\Lambda}{(d-2)(d-1)}r^2,
\end{align}
and
\begin{equation}
\Lambda=\frac{(d-2)(d-1)}{2L^2}, \quad d\Omega^2_{d-2}=d\chi_2^2+\prod_{i=2}^{d-2}\sin^2\chi_i\, d\chi_{i+1}^2,
\end{equation}
in which $q$ and $m$ are related to the electric charge $Q$ and the ADM mass $M$ of the BH, and $L$ is the cosmological radius. $M$ and $Q$
are expressed as
\begin{equation}
M=\frac{d-2}{16\pi}\omega_{d-2}m,\quad Q=\frac{\sqrt{2(d-2)(d-3)}}{8\pi}\omega_{d-2}q,\quad \omega_{d}=\frac{2\pi^{\frac{d+1}{2}}}{\Gamma(\frac{d+1}{2})},
\end{equation}
with $\omega_{d}$ being the volume of the unit $d$-sphere. The causal structure of a sub-extremal $d$-dimensional BH described by (\ref{metric}) admits three distinct horizons, where $r_-<\,r_+<\, r_c$ are the Cauchy, event and cosmological horizon radius, respectively. We denote the extremal electric charge of the BH as $Q_{\text{max}}$ at which the CH and event horizon coincide. The maximal cosmological constant is denoted as $\Lambda_{\text{max}}$ for each dimension,\footnote{e.g. for $d=4$, $\Lambda_\text{max}=2/9$, for $d=5$, $\Lambda_\text{max}=3\pi/4$ and for $d=6$, $\Lambda_{\text{max}}=\left(648\pi^2/25\right)^{\frac{1}{3}}$ provided that the black hole mass is set to $M=1$.} meaning that if $\Lambda>\Lambda_{\text{max}}$ holds then the spacetime would admit at most one horizon with positive radius, thus rendering our discussion irrelevant. To ensure the existence of  three distinct horizons, the cosmological constant must be restricted to $\Lambda<\Lambda_{\text{max}}$.
The surface gravity of each horizon is then
\begin{equation}
\label{surfGrav}
\kappa_h= \frac{1}{2}|f'(r_h)|\;\;,\; h\in\{-,+,c\}.
\end{equation}

The propagation of a neutral massless scalar field $\psi$ on a fixed $d$-dimensional RNdS background is described by the Klein-Gordon equation \cite{Berti:2009kk}. By expanding our field in terms of spherical harmonics
\begin{equation}
\psi=\sum_{lm}e^{-i\omega t}\frac{\phi(r)}{r^{\frac{d-2}{2}}}Y_{lm}(\chi)\label{eq3},
\end{equation}
we end up with the master equation
\begin{equation}
\frac{d^2\phi(r)}{dr_{\ast}^{2}}+(\omega^2-V_\text{eff})\phi(r)=0,\label{eq7}
\end{equation}
where
\begin{equation}
V_\text{eff}=f(r)\left(\frac{l(d+l-3)}{r^2}+\frac{(d-2)f'(r)}{2r}+\frac{f(r)(d-4)(d-2)}{4r^2}\right),\label{eq9}
\end{equation}
and $dr_{\ast}=\frac{dr}{f(r)}$ the tortoise coordinate. By imposing the boundary conditions
\begin{equation}\phi(r\rightarrow r_+)\sim e^{-i\omega  r_*},\,\,\,\,\,\,\,\,\,\,\,\,\,\,\,\phi(r\rightarrow r_c)\sim e^{i\omega {r}_*},\label{eq4}
\end{equation}
we select a discrete set of quasinormal frequencies called QNMs. Due to the similarity of characteristics of (\ref{eq9}) and the effective potential for odd (Regge-Wheeler \cite{Regge}) and even (Zerilli \cite{Zerilli,Zerilli2}) gravitational perturbations, the study of massless neutral scalar fields propagating on spherically symmetric backgrounds is a good proxy for more physically relevant gravitational field perturbations. 

As shown in Appendix \ref{appA}, for $d\geq 4$ the stability of the CH continues to be determined by (\ref{eq15}). The results shown in the following sections were obtained with the Mathematica package of \cite{Jansen:2017oag}, the asymptotic iteration method (AIM) \cite{AIM,Cho1}, and checked in various cases with a Wentzel-Kramers-Brillouin (WKB) approximation \cite{Iyer} and with a code developed based on the matrix method \cite{KaiLin1}.
\section{Dominant families of modes in higher-dimensional RNdS spacetime}
According to \cite{Cardoso1}, the region of interest in 4-dimensional RNdS, where violation of SCC may occur, lies close to extremality. There, the decay rate of perturbations in the exterior becomes comparable with the surface gravity of the CH $\kappa_-$ leading to $\beta>1/2$. We accumulate this result and scan the parameter space of higher-dimensional RNdS spacetimes for near-extremal parameters. By applying our numerics in the region of interest we discover three distinct families of modes.

The photon sphere (PS) is a spherical trapping region of space where gravity is strong enough that photons are forced to travel in unstable circular orbits around a BH. This region has a strong pull in the control of decay of perturbations and the QNMs with large frequencies. For instance, the decay timescale is related to the instability timescale of null geodesics near
the photon sphere. For asymptotically dS BHs, we find a family that can be traced back to the photon sphere and refer to them as PS modes. The dominant modes of this family are approached in the eikonal limit, where $l\rightarrow\infty$, and can be very well approximated with the WKB method (see Appendix \ref{appB}). For vanishing $\Lambda,\,Q$ they asymptote to the Schwarzschild BH QNMs in $d\geq 4$ dimensions. We find that $l=10$ provides a good approximation of the imaginary parts of the dominant modes which we depict in our plots with solid blue lines.

Our second family of modes, the BH de-Sitter (dS) family, corresponds to purely imaginary modes which can be very well approximated by the pure $d$-dimensional scalar dS QNMs \cite{Du:2004jt,LopezOrtega:2006my}:
\begin{align}
\label{dS1}
\omega_{\text{pure dS}}/\kappa_c^\text{dS}&=-i (l+2n),\\
\label{dS2}
\omega_{\text{pure dS}}/\kappa_c^\text{dS}&=-i (l+2n+d-1).
\end{align}
The dominant mode of this family ($n=0,\,l=1$) is almost identical to (\ref{dS1}) which we denote in our figures with red dashed lines. These modes are intriguing, in the sense that they have a surprisingly weak dependence on the BH charge and seem to be described by the surface gravity of $d$-dimensional dS $\kappa_c^{\text{dS}}=\sqrt{2\Lambda/(d-2)(d-1)}$ of
the cosmological horizon of pure $d$-dimensional dS space, as opposed
to that of the cosmological horizon in the RNdS BH under consideration.

Finally, as the CH approaches the event horizon, a new family of modes appears to dominate the dynamics. In the extremal limit of a $d$-dimensional RNdS BH the dominant ($n=l=0$) mode of this family approaches (see Appendix \ref{appC})
\begin{equation}
\label{NE}
\omega_\text{NE}=-i\kappa_{-}=-i\kappa_{+},
\end{equation}
where $\kappa_-,\,\kappa_+$ the surface gravity of the Cauchy and event horizon in $d$-dimensional RNdS spacetime. We call this family the near-extremal (NE) family of modes. Higher angular numbers $l$ admit larger (in absolute value) imaginary parts, thus rendered subdominant. In the asymptotically flat case, these modes seem to have been described analytically in the eikonal limit \cite{Zhang:2018jgj}.

\section{Strong Cosmic Censorship in higher-dimensional RNdS spacetime}

\begin{figure}[thbp]
\centering
\includegraphics[height=1.55in,width=2.15in]{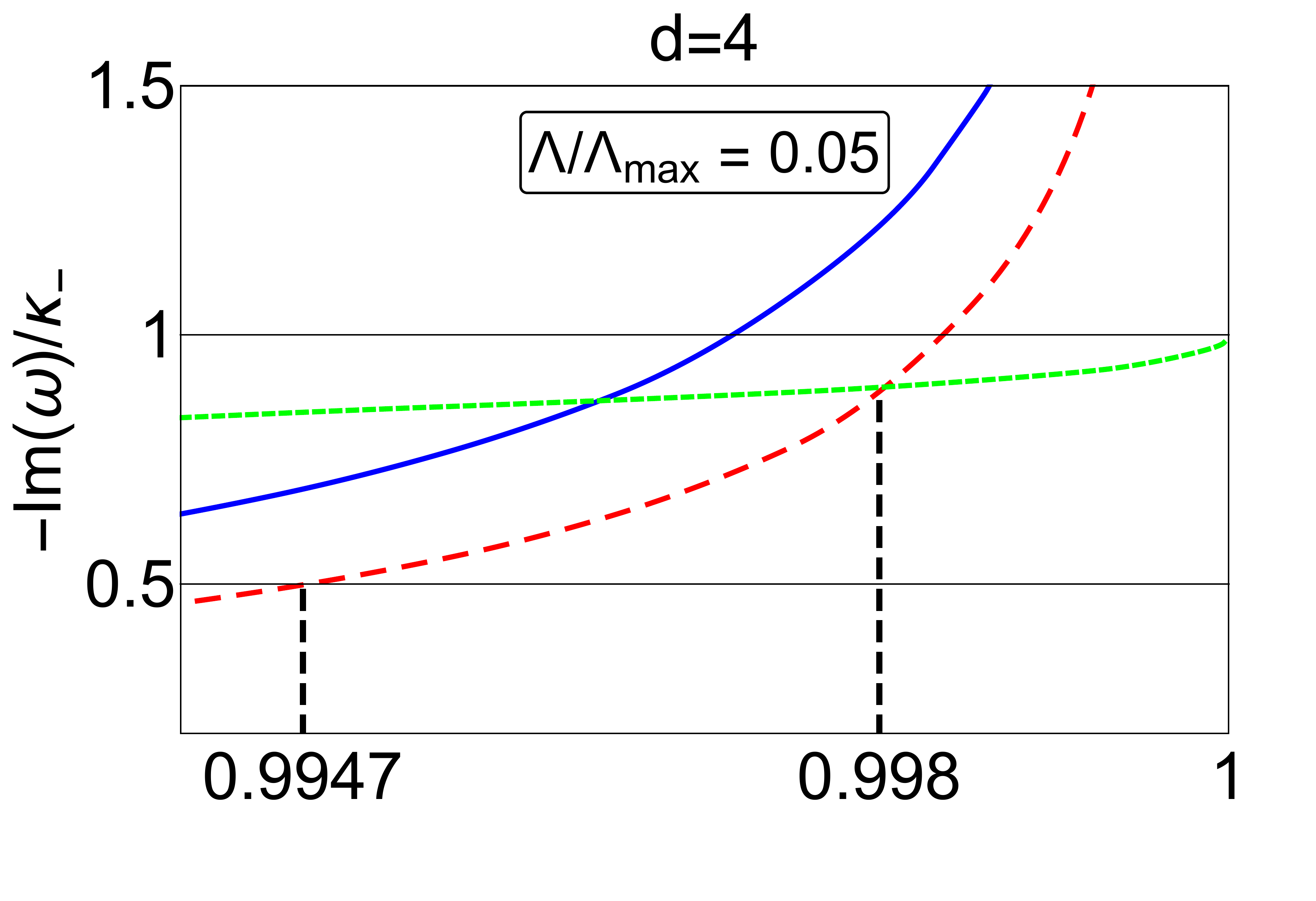}
\hskip -4.5ex
\includegraphics[height=1.55in,width=2.15in]{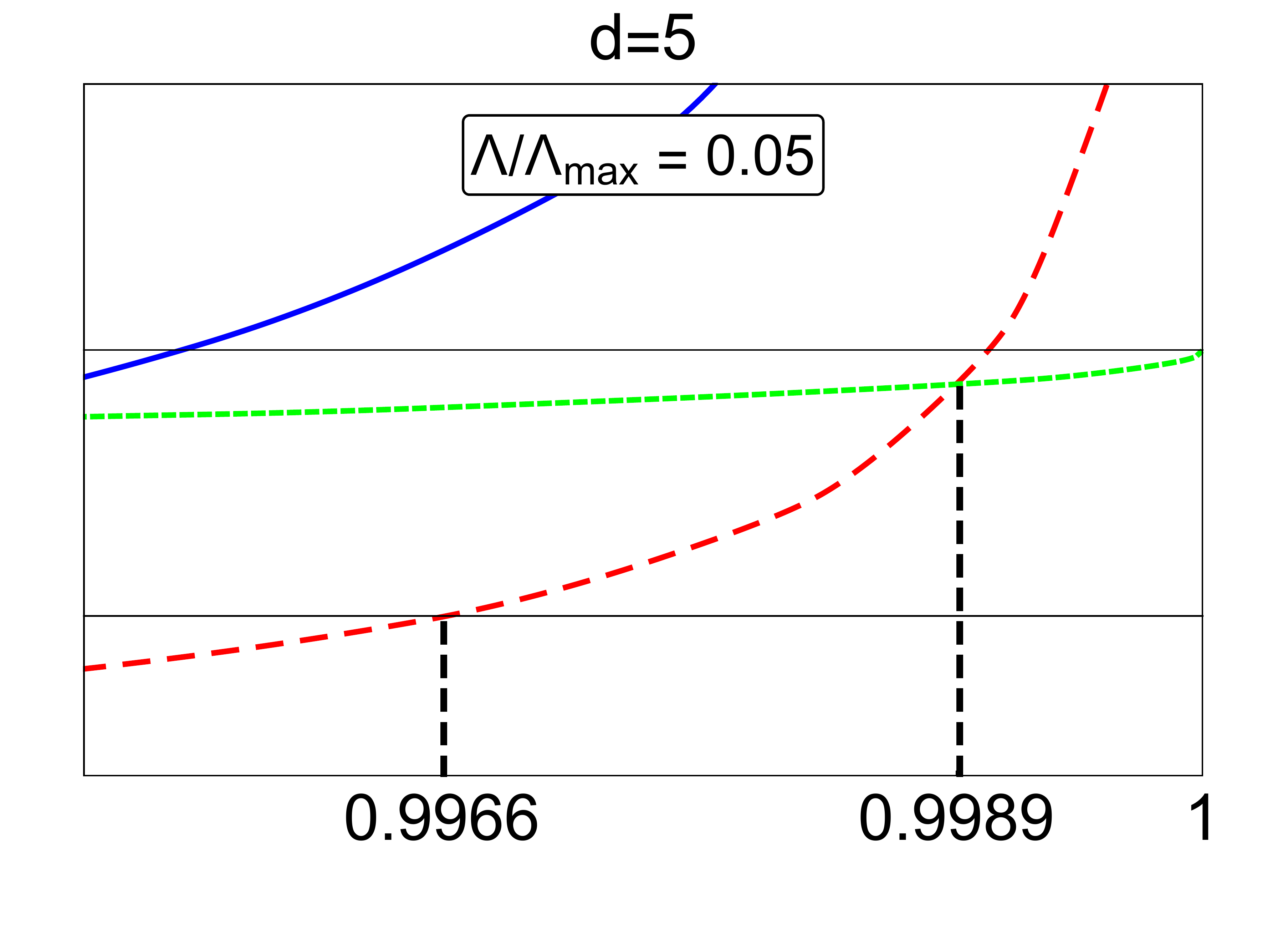}
\hskip -4.5ex
\includegraphics[height=1.55in,width=2.15in]{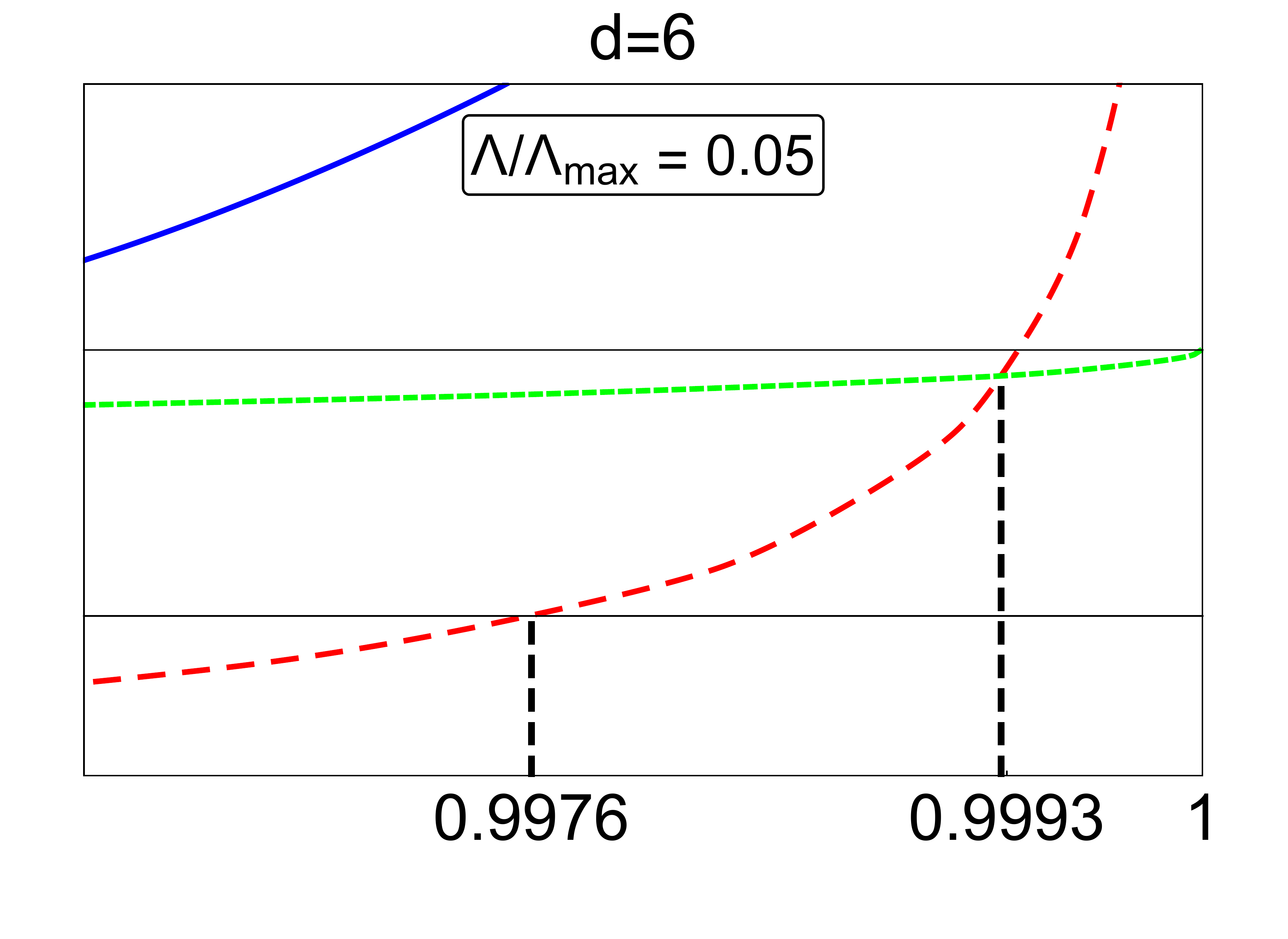}
\vskip -4ex
\includegraphics[height=1.55in,width=2.15in]{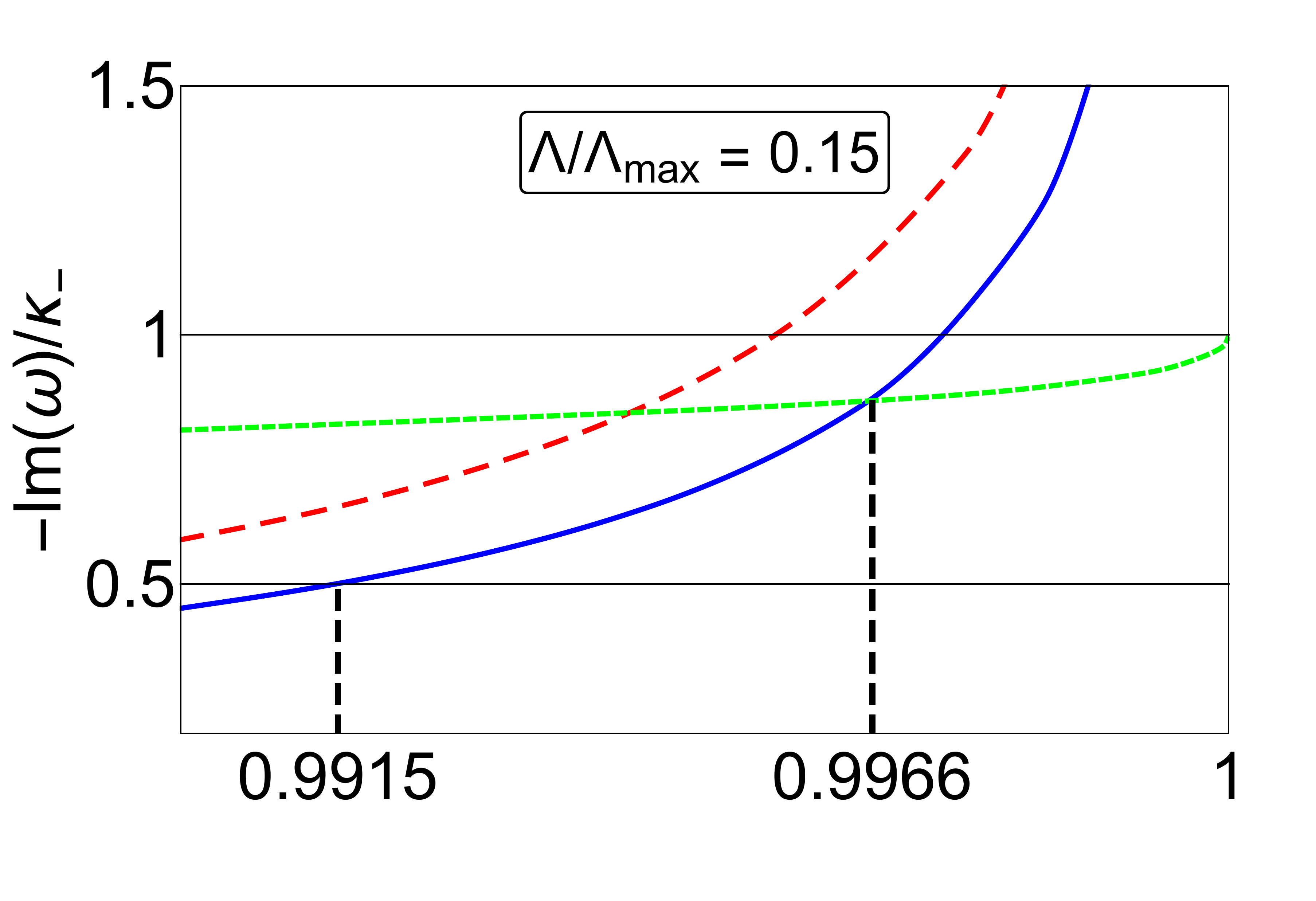}
\hskip -4.5ex
\includegraphics[height=1.55in,width=2.15in]{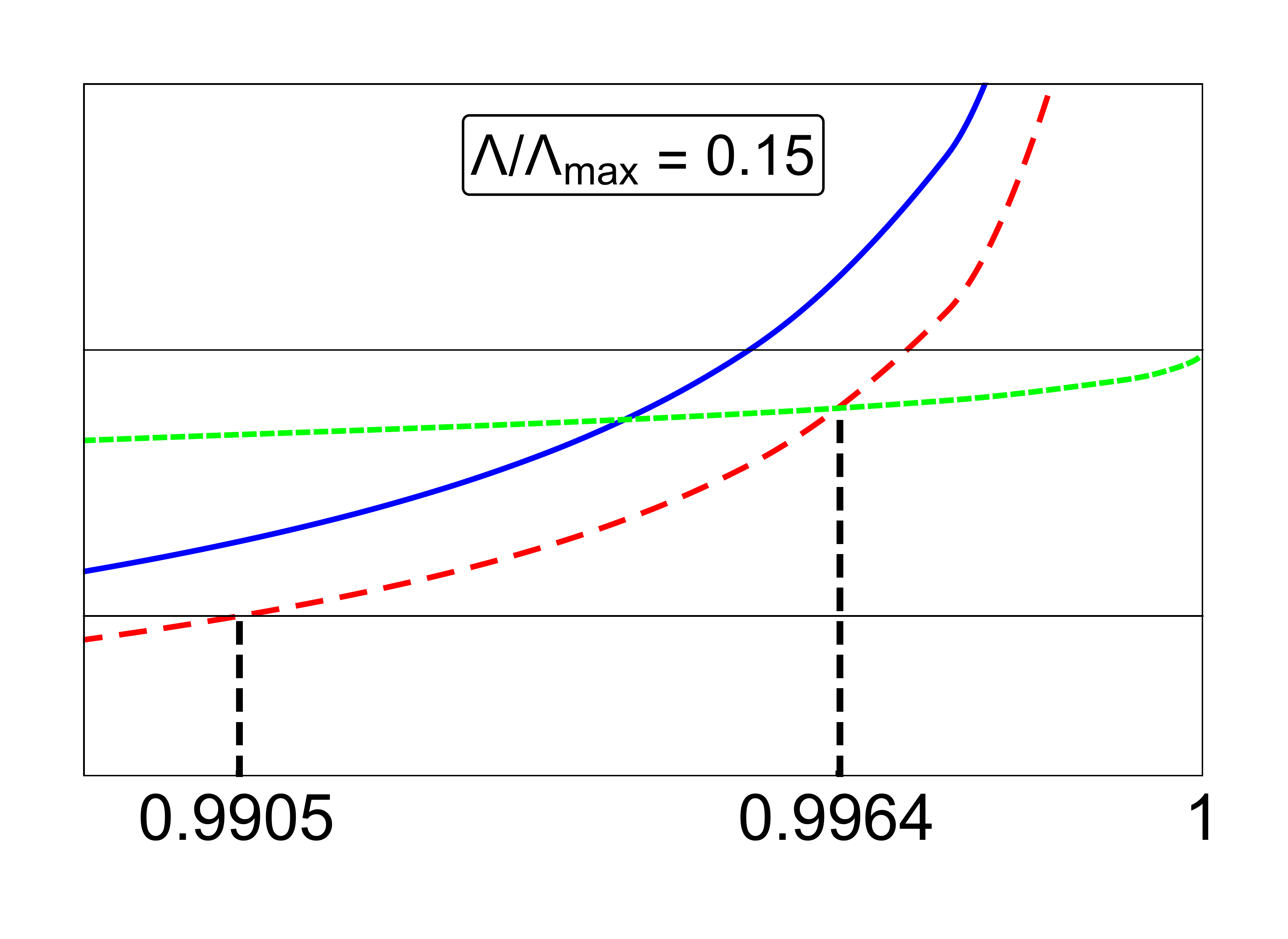}
\hskip -4.5ex
\includegraphics[height=1.55in,width=2.15in]{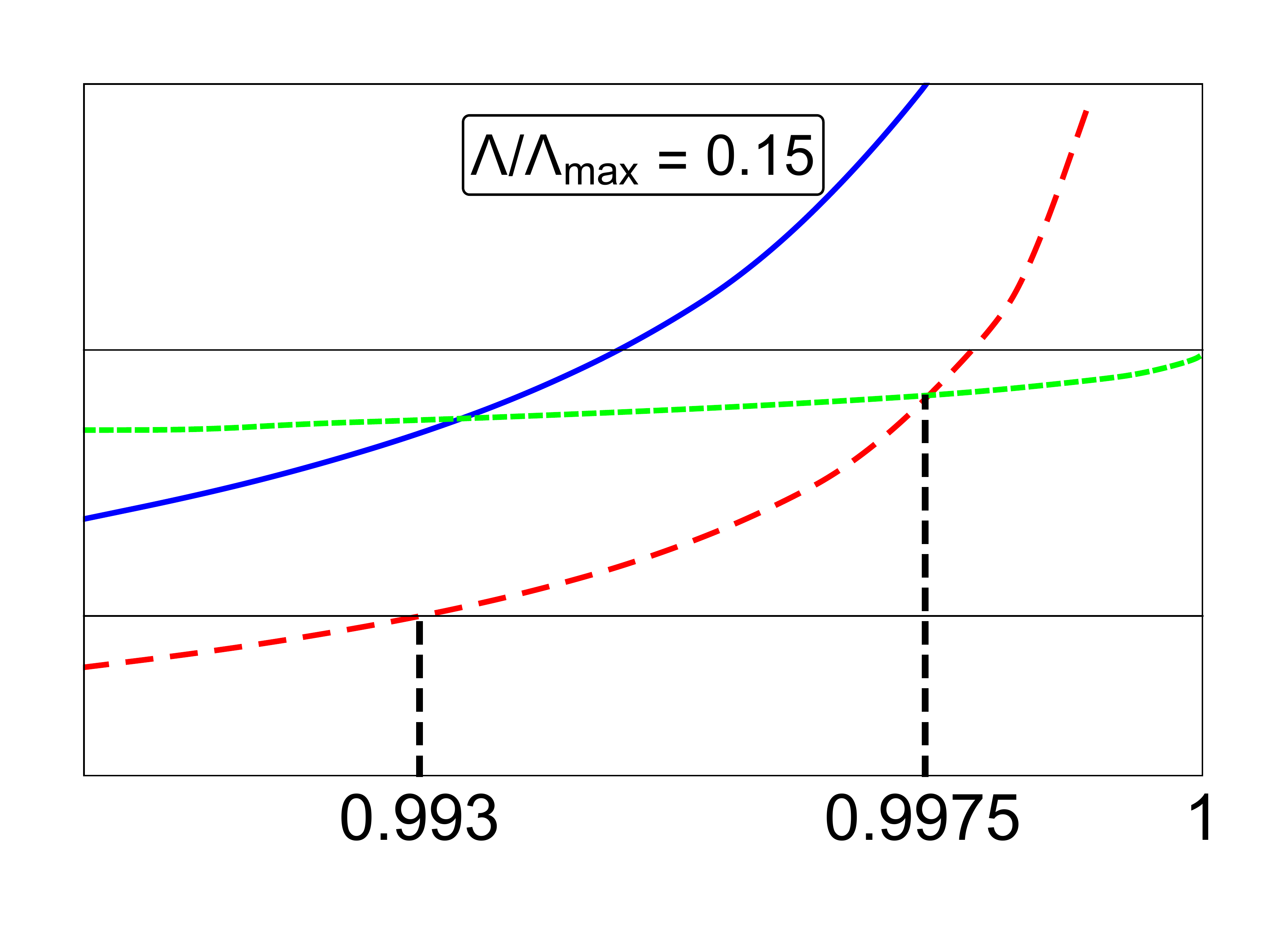}
\vskip -4ex
\includegraphics[height=1.55in,width=2.15in]{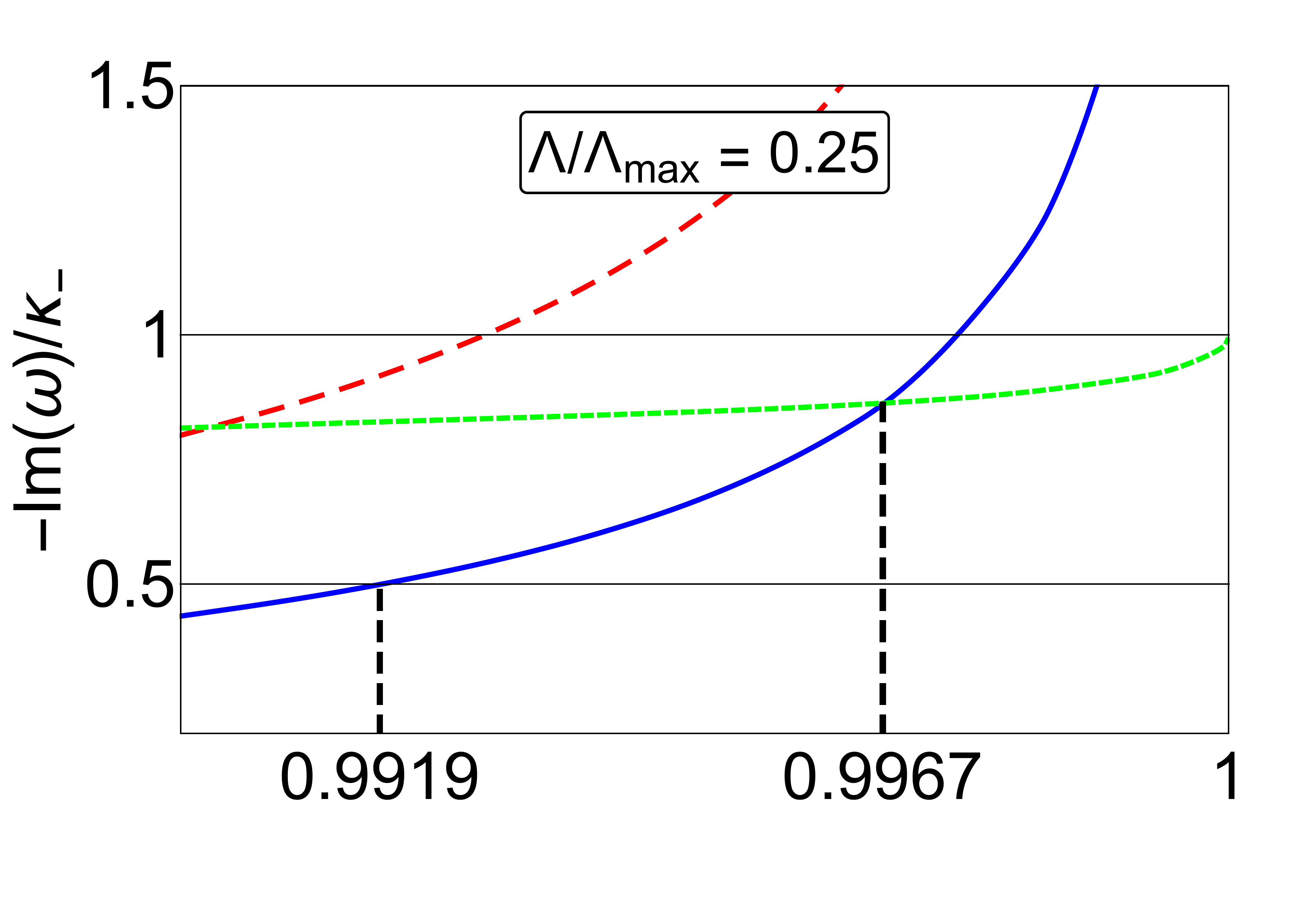}
\hskip -4.5ex
\includegraphics[height=1.55in,width=2.15in]{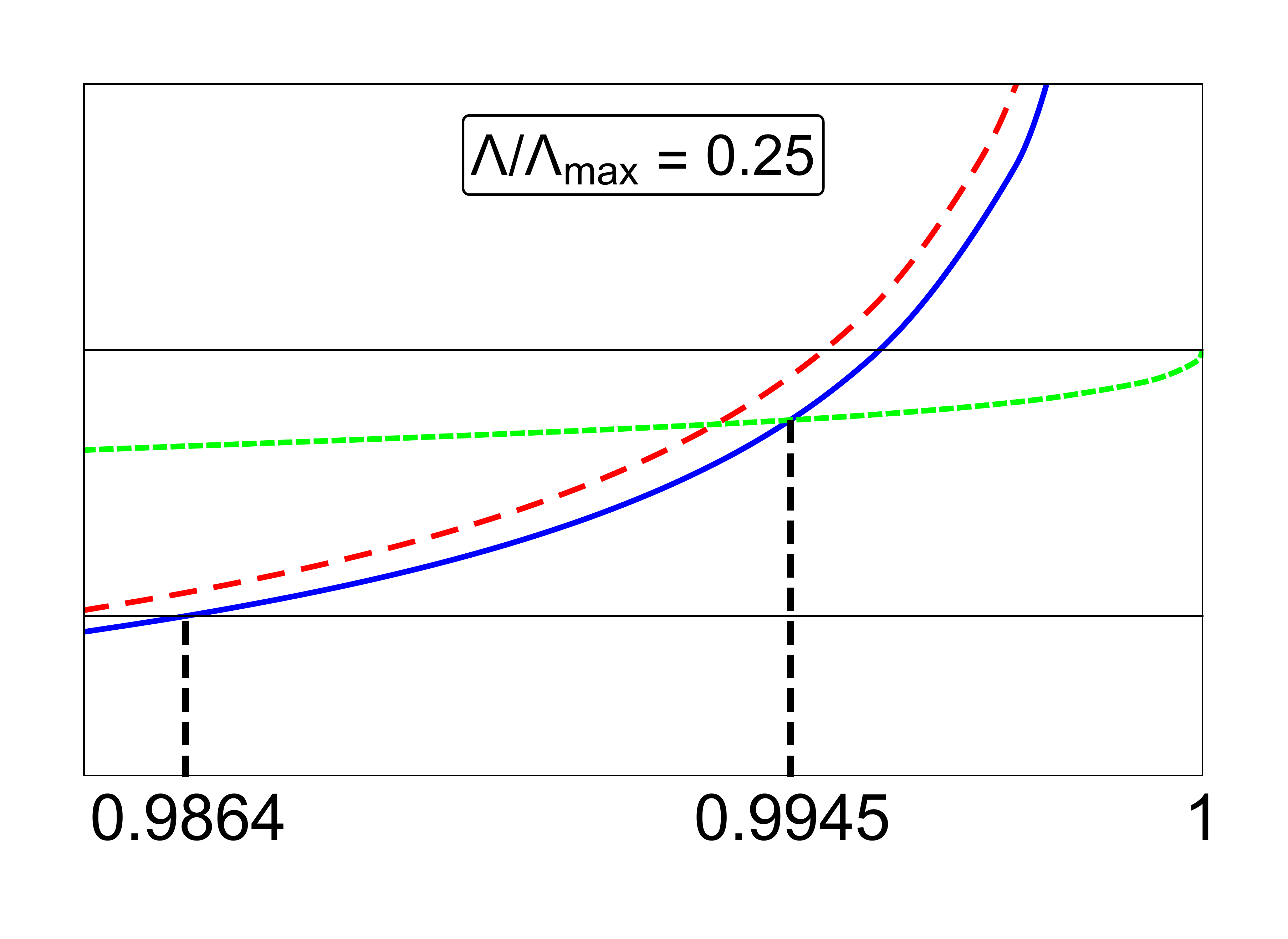}
\hskip -4.5ex
\includegraphics[height=1.55in,width=2.15in]{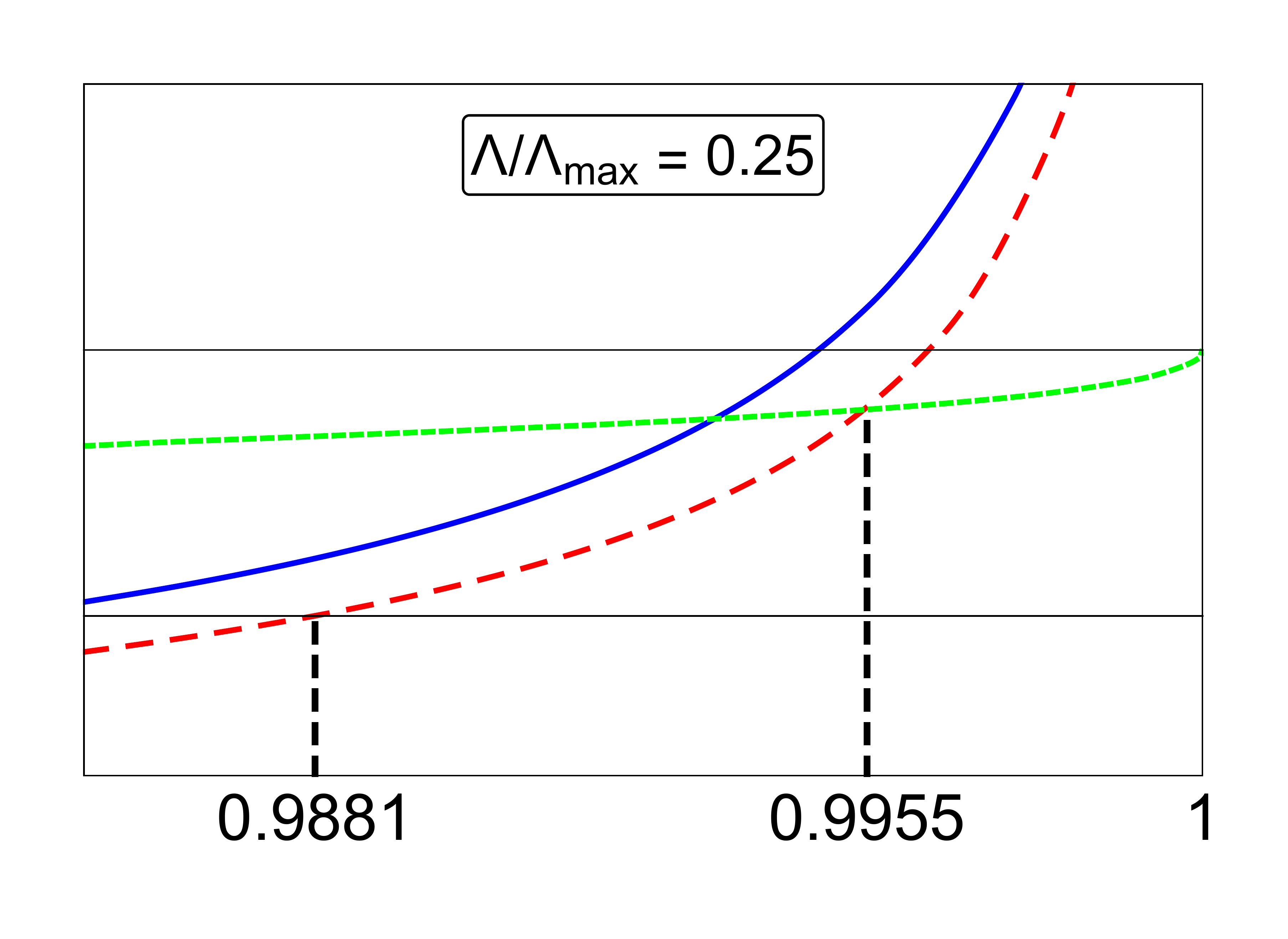}
\vskip -4ex
\includegraphics[height=1.55in,width=2.15in]{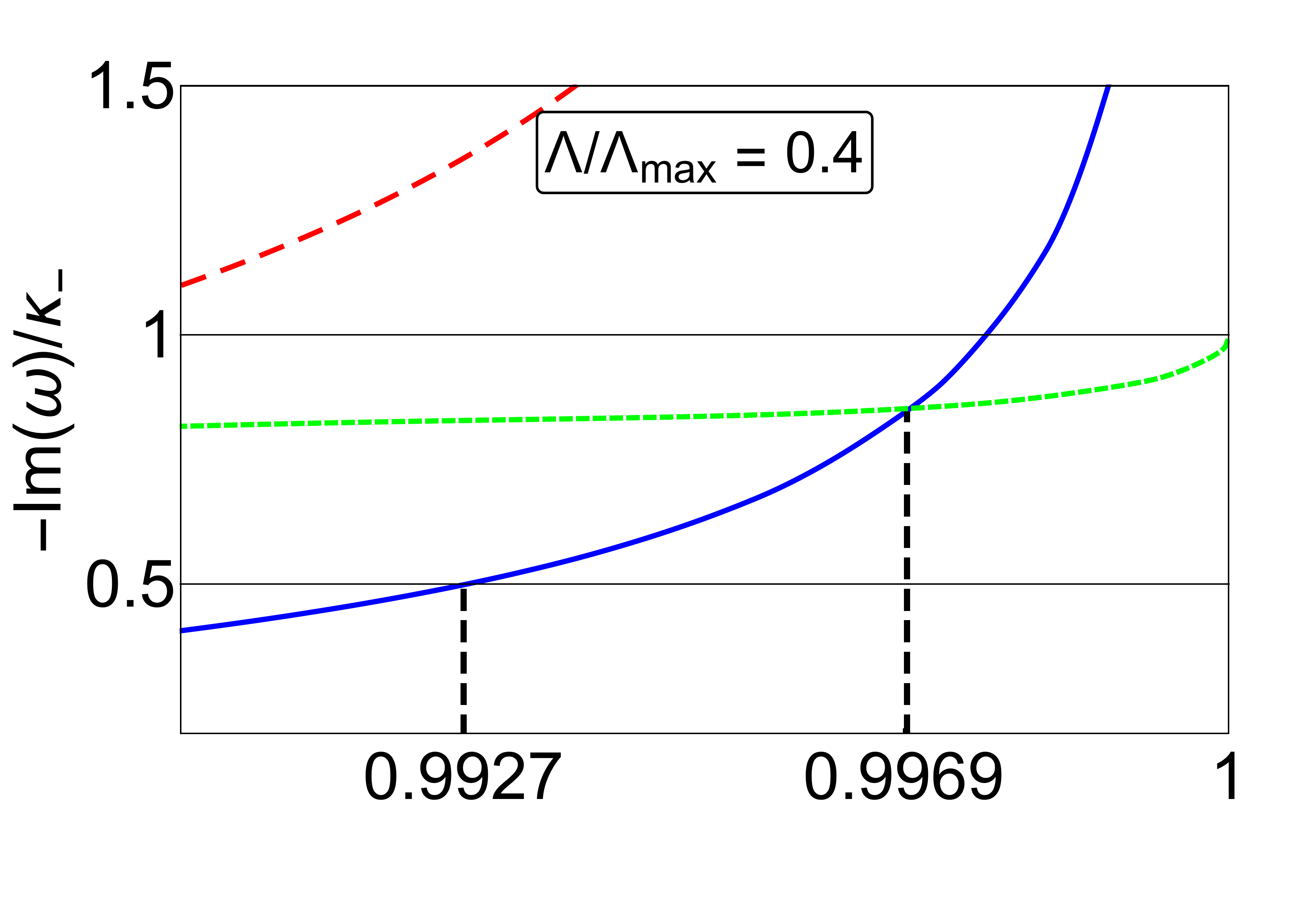}
\hskip -4.5ex
\includegraphics[height=1.55in,width=2.15in]{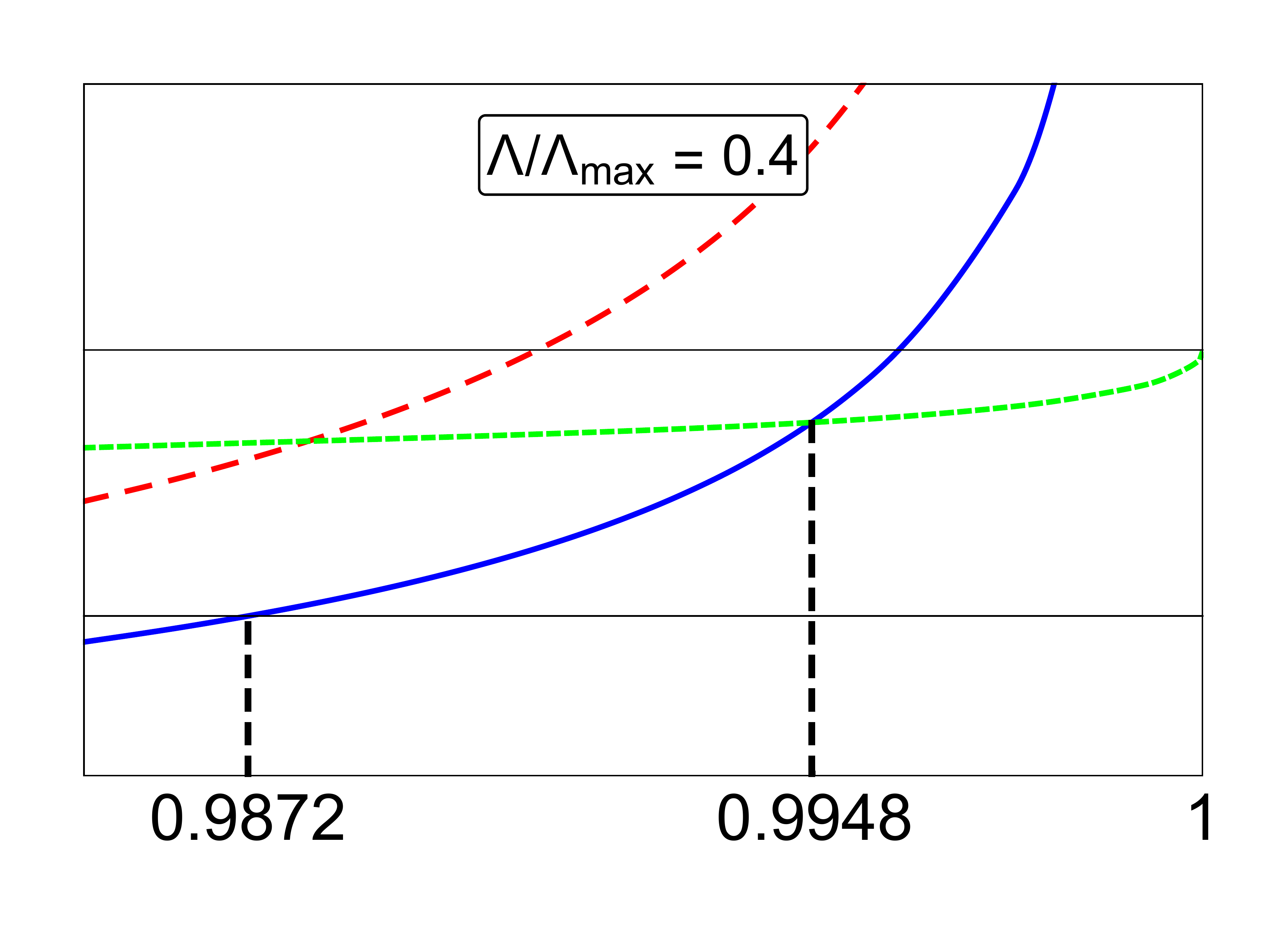}
\hskip -4.5ex
\includegraphics[height=1.55in,width=2.15in]{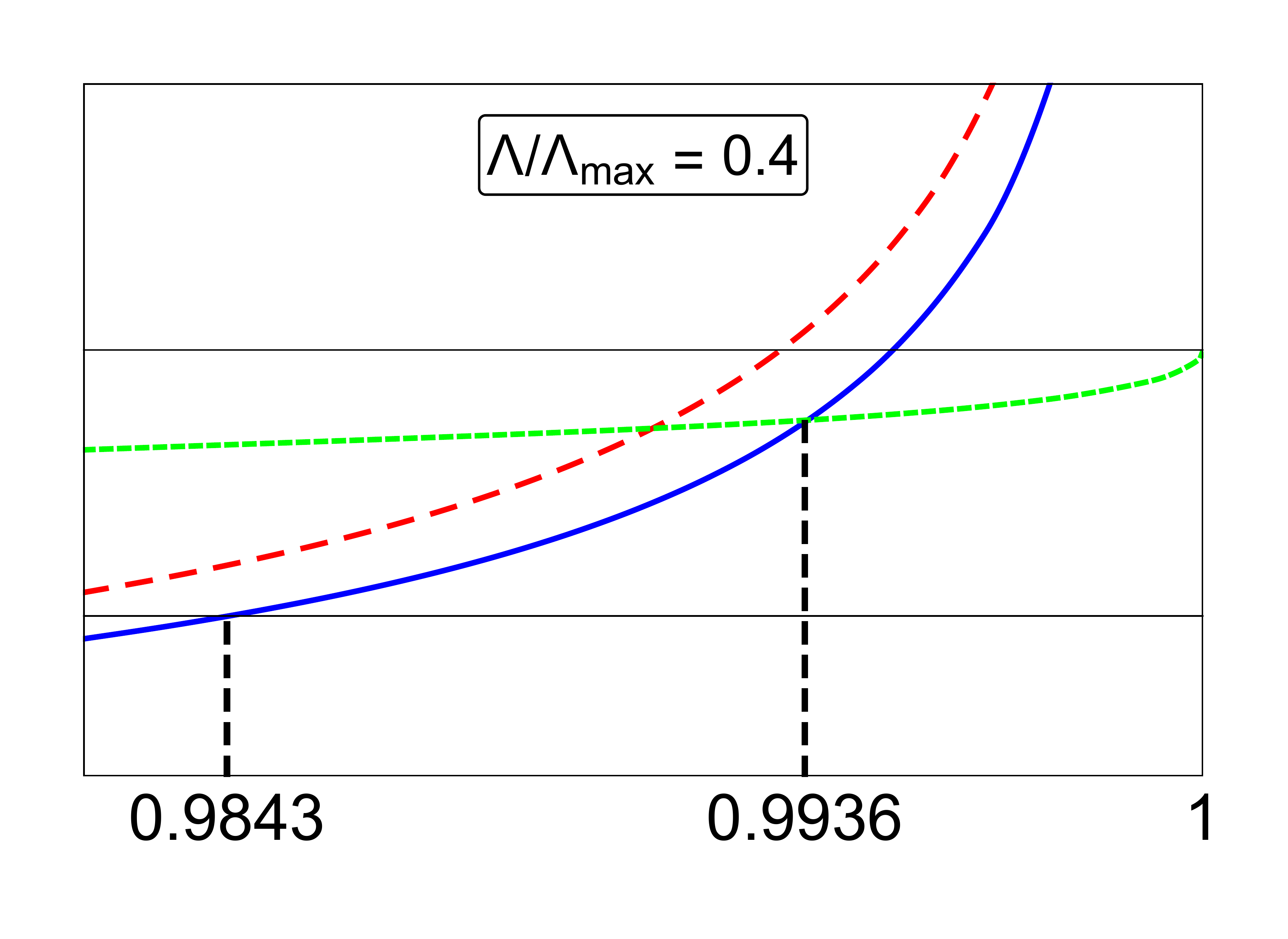}
\vskip -4ex
\includegraphics[height=1.54in,width=2.14in]{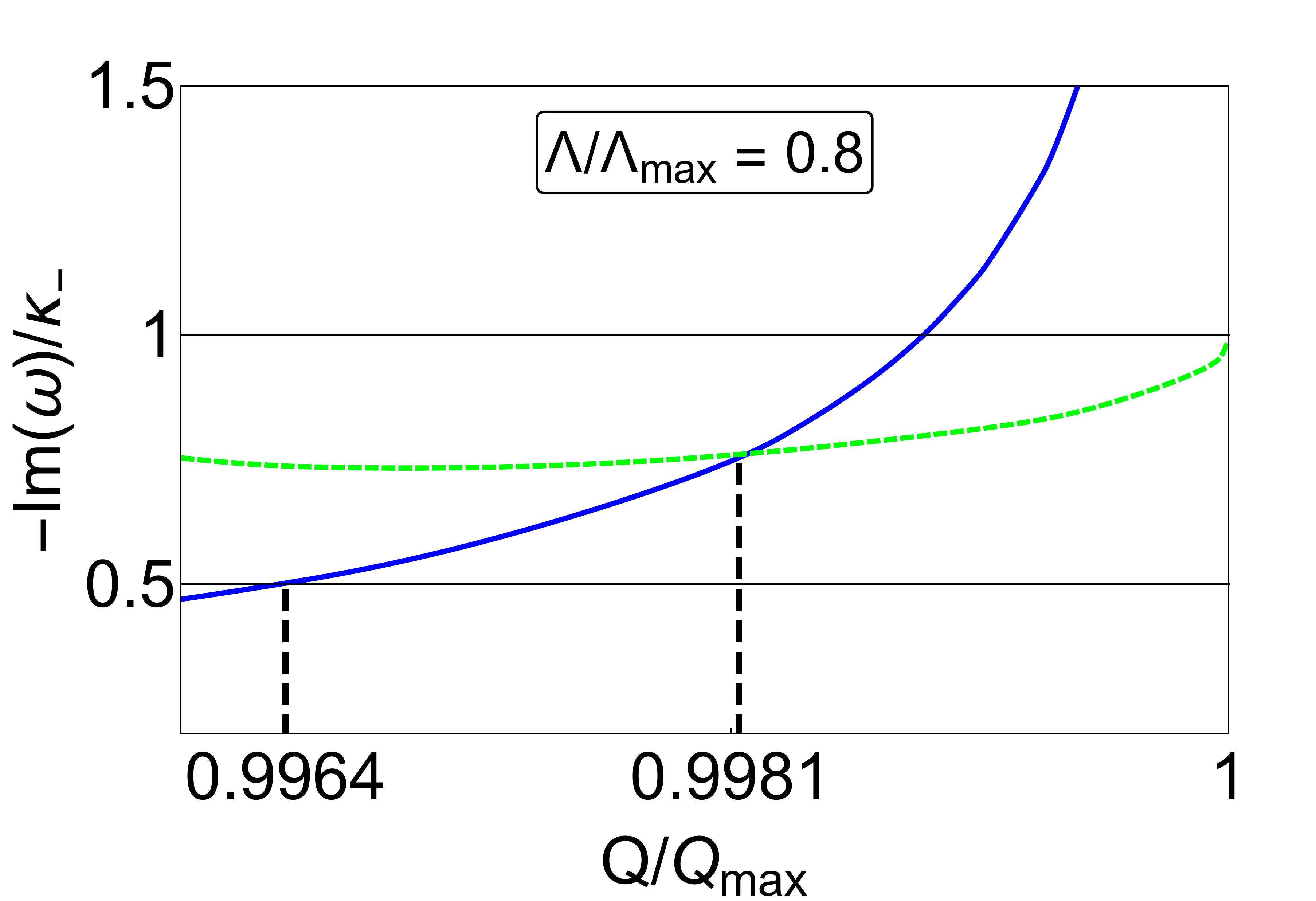}
\hskip -4.5ex
\includegraphics[height=1.54in,width=2.14in]{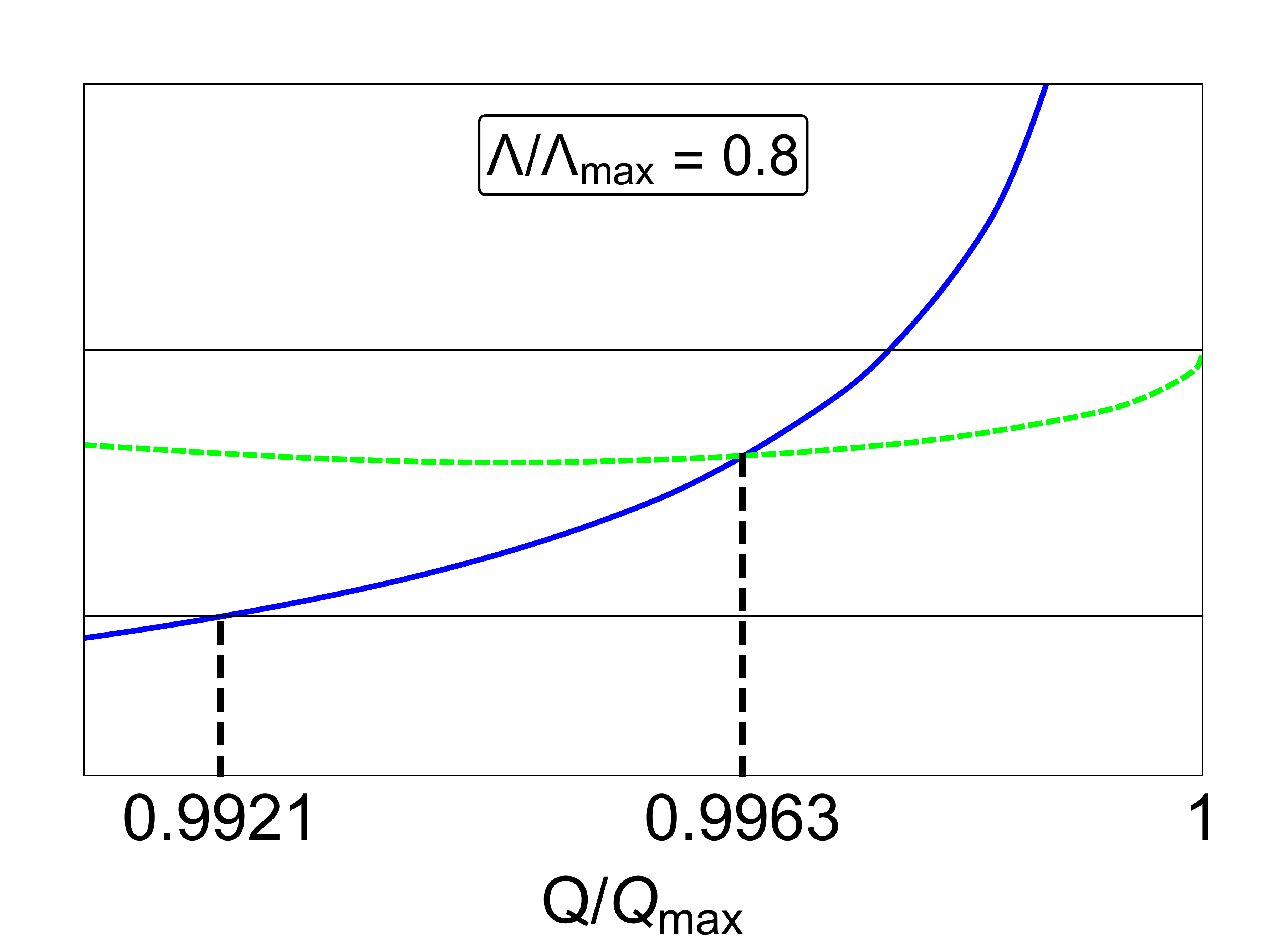}
\hskip -4.5ex
\includegraphics[height=1.54in,width=2.14in]{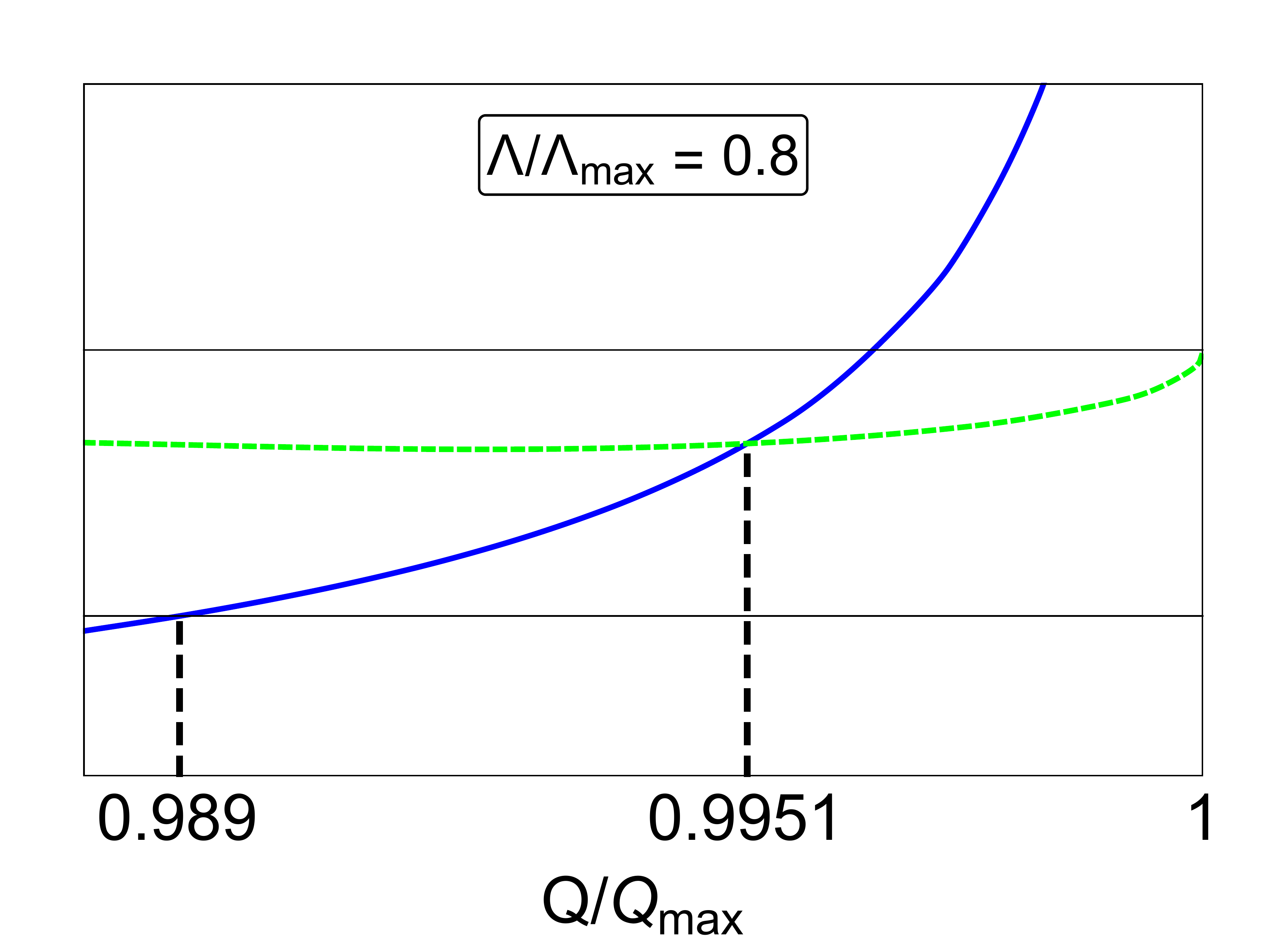}
\caption{Dominant modes of different families, showing the (nearly) dominant complex PS mode (blue, solid) at $l=10$, the dominant BH dS mode (red, dashed) at $l=1$ and the dominant NE mode (green, dashed) at $l=0$ for $d=4,\,5$ and $6$-dimensional RNdS spacetime with $M=1$. The two dashed vertical lines designate the points where $\beta=-\text{Im}(\omega)/\kappa_-=1/2$ and where the NE mode becomes dominant. The dS family in the final row of plots is too subdominant, thus lying outside of the region of interest.}
\label{beta}
\end{figure}

In Fig. \ref{beta} we depict the dominant modes of each of the previous families versus $\kappa_-$. We have chosen $d=4,\,5,$ and $6$-dimensional near-extremal RNdS BHs with various $\Lambda/\Lambda_\text{max}$. It is evident that for sufficiently "small" BHs\footnote{Usually we take $r_+/r_c$ to measure the size of small/big black hole, but in our discussion we compare black holes in different dimensions by fixing $\Lambda/\Lambda_{\text{max}}$ which has some connection with the size of black holes. It turns out that the value of $r_+/r_c$ would be notably influenced by $Q/Q_{\text{max}}$. On the other hand, if $Q/Q_{\text{max}}$ is fixed, one can find that the difference between $r_c$ and $r_+$ would increase with the decrease of $\Lambda/\Lambda_{\text{max}}$. For these reasons, the "small/large" black holes in this paper are only referred to black holes with small/large $\Lambda/\Lambda_{\text{max}}$.} (very small $\Lambda/\Lambda_{\text{max}}$), the increment of dimensions fortifies SCC for a larger region of the parameter space of $Q/Q_\text{max}$. On the other hand, for sufficiently "large" BHs (large $\Lambda/\Lambda_{\text{max}}$), the increment of dimensions work against the validity of SCC admitting violations for smaller $Q/Q_\text{max}$. To deepen into the understanding of this complex situation we denote the degree of difficulty of SCC violation with $d_4$ for $d=4$, $d_5$ for $d=5$ and $d_6$ for $d=6$. For example, for the case of $\Lambda/\Lambda_\text{max}=0.05$, the degree of difficulty of SCC violation follows $d_6>d_5>d_4$, meaning that 6-dimensional RNdS BHs require the highest BH charge to be violated. The second hardest BH to be violated is the 5-dimensional and, finally, the easiest to be violated is the 4-dimensional.

In the "intermediate" region, where $\Lambda/\Lambda_\text{max}$ is neither too small nor too large, the picture becomes obscured by the delicate interplay of the QNMs of the dominant PS and dS family. To that end, we have depicted two interesting cases. In the first case, for $\Lambda/\Lambda_\text{max}=0.15$, the degree of difficulty to violate SCC follows $d_6>d_4>d_5$, while in the second case for $\Lambda/\Lambda_\text{max}=0.25$ we have $d_4>d_6>d_5$. This perplex picture appears due to the opposite behavior that the dominant PS and dS family possess. As shown in Fig. \ref{beta}, higher dimensions oblige $\beta_\text{PS}$, as it gets determined by the dominant modes of the PS family ($l=10$), to move upwards in the plots, thus becoming subdominant, while $\beta_\text{dS}$, as it gets determined by the dominant modes of the dS family ($l=1$), moves downwards. On the other hand, the increment of $\Lambda/\Lambda_{\text{max}}$ has the opposite effect on these families as expected by \cite{Cardoso1}. It is easy to realize (see the pattern in Fig. \ref{beta}) that the inclusion of even higher than 6 dimensions will make the picture of "intermediate" and "large" BHs even richer and much more perplex\footnote{E.g. for $\Lambda/\Lambda_{\text{max}}=0.4$ we can see that if $d=7$ or 8 where to be included, then the dS family would eventually dominate for such dimensions, thus changing the picture into a richer version of $\Lambda/\Lambda_{\text{max}}=0.15$ or $0.25$.}. The only solid case is the one for "small" BHs. There, $\beta$ is essentially (for the largest part of the parameter space) determined by the dominant modes of the dS family which will become even more dominant for increasing dimensions if no new families or instabilities occur in $d>6$ dimensions.\footnote{It is natural to question whether more slowly decaying modes might appear in the dimensions considered. For this purpose, we have used calculations of very high accuracy to rule out the possibility of lost dominant modes. This means that, if more families do exist, they should be subdominant thus are irrelevant for SCC.}

To distinguish between "small", "intermediate" and "large" BHs, we scan thoroughly the parameter space of $d=4,\,5$ and 6-dimensional RNdS BHs to find critical values of $\Lambda/\Lambda_{\text{max}}$ where different violation configurations are introduced. We find 3 critical values which divide the range of $0<\Lambda/\Lambda_{\text{max}}\leq 1$ into 4 regions. In Table \ref{table2}, we summarize the division of our parameter space into the regions of interest and display the degree of difficulty of SCC violation at each region. We realize that region I corresponds to "large" BHs, regions II and III correspond to "intermediate" BHs and, finally, region IV corresponds to "small" BHs. These regions can be directly seen in Fig. \ref{beta} and arise due to the existence and competition between the dominant PS and dS family, as discussed above.

\begin{table}[!htbp]
\centering
\begin{tabular}{||c|c||}
\hline
parameter regions  & degree of difficulty of SCC violation\\
\hline
\text{\bf I.\,\,}$\Lambda/\Lambda_{\text{max}}\gtrsim 0.279$ & $d_4>d_5>d_6$ \\
\hline
\text{\bf II.\,\,}$0.179\lesssim\Lambda/\Lambda_{\text{max}}\lesssim 0.279$ & $d_4>d_6>d_5$ \\
\hline
\text{\bf III.\,\,}$0.135\lesssim\Lambda/\Lambda_{\text{max}}\lesssim 0.179$ & $d_6>d_4>d_5$\\
\hline
\text{\bf IV.\,\,\,}$\Lambda/\Lambda_{\text{max}}\lesssim 0.135$ & $d_6>d_5>d_4$ \\
\hline
\end{tabular}
\caption{Comparison of $\Lambda/\Lambda_{\text{max}}$ with respect to the degree of difficulty of SCC violation in $d=4,\,5$ and 6-dimensional RNdS BHs.
\label{table2}}
\end{table}

In any case, we clearly see that $\beta>1/2$ above some value of the BH charge, no matter the choice of the cosmological constant. This leads to CHs which upon scalar perturbations maintain enough regularity for the scalar field (and thus the metric) to be extendible past it, resulting to a potential violation of SCC. Moreover, if it was up to the PS and dS family, $\beta$ would always diverge at extremality. However, the dominant modes of the NE family ($l=n=0$) will always take over to keep $\beta$ below 1.

\section{Conclusions and Discussions}
The studies of \cite{Costa3,Hintz1,Hintz3} indicate that the stability of the CH in asymptotically dS spacetimes is governed by $\beta$ defined in (\ref{eq15}). Subsequently, the results of \cite{Cardoso1} indicate a potential failure of determinism in General Relativity when near-extremal 4-dimensional RNdS BHs are considered. Under massless neutral scalar perturbations, the CH might seem singular, due to the blow-up of curvature components, but maintain enough regularity as to allow the field equations to be extended beyond a region where the evolution of gravitation is classically determined in a highly non-unique manner.

Here, we extend our study to higher-dimensional RNdS BHs and find that the same picture occurs when scalar fields are considered. We have proven that (\ref{eq15}) remains unchanged for $d$-dimensions. By inferring to $d=4,\,5$ and 6-dimensional RNdS BHs we realize that the introduction of higher dimensions will fortify SCC for sufficiently "small" BHs ($\Lambda/\Lambda_{\text{max}}\lesssim 0.135$), by the introduction of higher BH charges beyond which $\beta>1/2$. Moreover, we observe that "intermediate" and "large" BHs ($\Lambda/\Lambda_{\text{max}}\gtrsim 0.135$) possess a much more complex picture with some dimensions being preferred over others to fortify SCC. This perplexity arises due to the delicate competition of the PS and dS family of modes. Even though for "large" BHs we see that the preferred dimension to fortify SCC, with higher $Q/Q_\text{max}$ beyond which $\beta>1/2$, is $d=4$, we understand that the introduction of even higher than 6 dimensions will eventually change the picture due to the behavior of the dS family demonstrated in Fig. \ref{beta}, if no instabilities occur in our region of interest \cite{Konoplya1}.

In any case, we can always find a region in the parameter space of the higher-dimensional RNdS BHs in study for which $\beta$ exceeds $1/2$, but still not exceeding unity\footnote{$\beta>1$ would correspond to extensions of the scalar field in $C^1$ at the CH, thus the coupling to gravity should lead to the existence of solutions with bounded curvature.}. This still leaves as with CHs which upon perturbations might seem singular, due to the blow-up of curvature components, but that doesn't imply the breakdown of Einstein's field equations \cite{Klainerman:2012wt} nor the destruction of macroscopic observers \cite{Ori} at the CH.

It is important to mention that SCC in higher-dimensional RNdS spacetime was also discussed in \cite{Rahman}, with a wishful premise that the large $l$ mode always dominates, i.e., the value of $\beta$ always decreases monotonously with the increase of angular number $l$. However, this is not the case, as we have seen in Fig. \ref{beta}, due to the existence of three different families of modes. In fact, the existence of more families highly affects $\beta$ according to the choice of our cosmological constant and the dimensions of our spacetime, thus rendering the study in \cite{Rahman} incomplete.

{\bf Note added:} An updated version of \cite{Rahman} was published recently. In the new modified version, their improved results are in agreement with ours, thus confirming our findings.

\section*{Acknowledgements}
We are grateful to Zhen Zhong for his helpful discussions in the early stage of this project. This work is partially supported by NSFC with Grant No.11475179, No. 11675015, No. 11775022, and No. 11875095 as well as by FWO-Vlaanderen through the project G020714N, G044016N, and G006918N. BW acknowledges the support by NSFC with Grant No.11575109. HZ is an individual FWO Fellow supported by 12G3515N and by the Vrije Universiteit Brussel through the Strategic Research Program High-Energy Physics. KD acknowledges financial support provided under the European Union's H2020 ERC Consolidator Grant "Matter and strong-field gravity: New frontiers in Einstein's theory" grant agreement no. MaGRaTh--646597. KD, also, acknowledges networking support by the GWverse COST Action CA16104, "Black holes, gravitational waves and fundamental physics".

\begin{appendix}
\section{The definition of $\beta$ in higher-dimensional spherically symmetric spacetimes}\label{appA}

In \cite{Cardoso2,Destounis1} a justification of searching for $\beta>1/2$ was provided, leading to potential violation of SCC in 4-dimensional RNdS BHs under charged scalar and fermionic perturbations. Here, we prove that the same holds for neutral massless scalar perturbations in $d$-dimensional RNdS spacetime. To determine the regularity of the metric up to the CH we study the regularity of QNMs at the CH. To do so, we change to outgoing Eddington-Finkelstein coordinates which are regular there.  The outgoing Eddington-Finkelstein coordinates are obtained by replacing $t$ with $u=t-r_*$ in (\ref{dspace}) to get
\begin{equation}
\label{outgoing}
ds^2=-f(r)du^2-2du dr +r^2d\Omega_{d-2}^2.
\end{equation}
By expanding the Klein-Gordon equation
\begin{equation}
\Box\psi=0,
\end{equation}
we get $P\psi=0$ where the operator $P$ reads
\begin{equation}
\label{int1}
P\psi=-2\partial_u\partial_r\psi-\frac{d-2}{r}\partial_u\psi+\frac{1}{r^{d-2}}\partial_r\left(f r^{d-2}\partial_r\psi\right)+\frac{\Delta_{\Omega_{d-2}}}{r^{d-2}}\psi,
\end{equation}
where $\Delta_{\Omega_{d-2}}$ the Laplace-Beltrami operator \cite{Berti:2009kk}. By acting on mode solutions of the form $\psi\sim e^{-i\omega u} \phi$ we obtain
\begin{equation}
\label{int}
f P\psi=2i\omega f\partial_r\phi +\frac{i\omega(d-2)}{r^{d-2}}f\phi+\frac{1}{r^{d-2}}f\partial_r\left(f r^{d-2}\partial_r\phi\right)+\frac{\Delta_{\Omega_{d-2}}}{r^{d-2}}f\phi.
\end{equation}
It can be shown that the mode solutions of (\ref{int}) are conormal at $r=r_-$, meaning that they grow at the same rate $|r-r_-|^\lambda$. Thus, if $\phi\sim|r-r_-|^\lambda$ then the second and last term have higher regularity than the rest, since $f \sim |r-r_-|$ near the CH. This means that these terms can be neglected, which leads to a regular-singular ordinary differential equation near $r=r_-$ of the form $\tilde{P}\phi=fP\phi=0$ with the operator
\begin{equation}
\label{operator}
\tilde{P}=2i\omega f\partial_r+\left(f\partial_r\right)^2.
\end{equation}
It is convenient to use $f$ as a radial coordinate instead of $r$, so $\partial_r=f^\prime\partial_f=f^\prime(r_-)\partial_f$ near the CH modulo irrelevant terms. Moreover, the surface gravity at the CH is $\kappa_-=-f^\prime(r_-)/2$ so $f\partial_r=-2\kappa_-(f\partial_f)$. Thus, (\ref{operator}) becomes
\begin{equation}
\label{indicial}
\frac{\tilde{P}}{4\kappa_-^2}=(f\partial_f)^2-\frac{i\omega}{\kappa_-}\left(f\partial_f\right)=f\partial_f\left(f\partial_f-\frac{i\omega}{\kappa_-}\right).
\end{equation}
It remains to calculate the allowed growth rates $\lambda$, i.e. indicial roots of the differential operator (\ref{indicial}). Acting with $|f|^\lambda$ we get
\begin{equation}
\label{polynomial}
\frac{{P}}{4\kappa_-^2}|f|^\lambda=\lambda\left(\lambda-\frac{i\omega}{\kappa_-}\right)|f|^\lambda.
\end{equation}
The indicial roots are the roots of the quadratic polynomial (\ref{polynomial}), namely
\begin{equation}
\lambda_1=0,\,\,\,\,\,\,\,\,\,\,\,\,\,\lambda_2=\frac{i\omega}{\kappa_-}.
\end{equation}
The root $\lambda_1=0$ corresponds to mode solutions which are approximately constant, i.e. remain smooth at the CH and are irrelevant for SCC, while $\lambda_2$ corresponds to asymptotics
\begin{equation}
|f|^{\lambda_2}\sim|r-r_-|^{\frac{i\omega}{\kappa_-}}.
\end{equation}
If we consider QNMs of the form $\omega=\omega_R-i\omega_I$ then
\begin{equation}
|f|^{\lambda_2}\sim|r-r_-|^\frac{\omega_I}{\kappa_-}|r-r_-|^\frac{i\omega_R}{\kappa_-}.
\end{equation}
The second factor is purely oscillatory, so the only relevant factor for SCC is $|r-r_-|^\frac{\alpha}{\kappa_-}$ with $\alpha:=-\text{Im}({\omega})$ the spectral gap defined in \cite{Cardoso1}. This function lies in the Sobolev space $H^s$ for all $s<\frac{1}{2}+\frac{\alpha}{\kappa_-}$.

Since we are considering scalar fields, we require locally
square integrable gradient of the scalar field at the CH\footnote{The energy-momentum tensor for scalar fields is $T_{\mu\nu}\sim \partial_\mu\psi\partial_\nu\psi$ which with proper manipulation can be bounded by squares of the gradient of $\psi$.}, i.e., the mode solutions should belong to the Sobolev space $H^1_\text{loc}$ for the Einstein's field equations to be satisfied weakly at the CH. This justifies our search for $\beta=-\text{Im}(\omega)/\kappa_->1/2$.
\section{WKB approximation of the dominant photon sphere modes}\label{appB}
The WKB method can provide accurate approximation of QNMs in the eikonal limit. The QNMs of BHs in the eikonical limit under massless scalar perturbations are related to the Lyapunov exponent $\lambda$ of the null unstable geodesic, which is inversely-proportional to the instability timescale associated with the geodesic motion of null particles near the photon sphere. For $d$-dimensions we have \cite{Cardoso4}
\begin{equation}
\begin{split}
\omega_\text{WKB}&=l\sqrt{\frac{f(r_s)}{r_s^2}}-i\left(n+\frac{1}{2}\right)\sqrt{-\frac{1}{2}\frac{r_s^2}{f(r_s)}\left(\frac{d^2}{dr_\ast^2}\frac{f(r)}{r^2}\right)_{r_s}}\\
&=\Omega_c l-i\left(n+\frac{1}{2}\right)\left|\lambda\right|,
\end{split}
\end{equation}
where $r_s$ is the radius of the null circular geodesic, and $\Omega_c$ the coordinate angular velocity of the geodesic. By focusing on the modes with overtone number $n=0$, we have $\beta=\left|\lambda\right|/{2\kappa_-}$ for the dominant PS modes at the eikonical limit.

\begin{table}[!htbp]
\centering
\begin{tabular}{||c|c|c|c||}
\hline
$\beta_\text{PS}$& $d=4$& $d=5$&$d=6$\\
\hline
$\beta_\text{WKB}\,\,\,\,\,\,(l\rightarrow\infty)$&0.328192 & 0.687518 & 0.775677\\
\hline
$\beta_\text{AIM}\,\,\,\,\,\,\,\,(l=10)$&0.328304& 0.689089 & 0.778164\\
\hline
$\beta_\text{spectral} \,(l=10)$& 0.328304&0.689089&0.778164\\
\hline
\end{tabular}
\caption{Comparison of $\beta_\text{PS}$ obtained with WKB, AIM and a spectral method for a RNdS BH with $M=\frac{1}{3\sqrt{2}}$, $\Lambda=3$ and $Q/Q_\text{max}=0.992$.}\label{table1}
\end{table}

In Table \ref{table1}, we compare the value of $\beta$ obtained by AIM and the spectral method \cite{Jansen:2017oag} at $l=10$ and the value evaluated by the WKB method at large $l$ for the same BH parameters. We observe that the difference between $\beta_\text{WKB}$, $\beta_\text{spectral}$ and $\beta_\text{AIM}$ is very small, meaning that the choice of $l=10$ in our numerics can be regarded as a good approximation of $\beta_\text{WKB}$ of the dominant PS modes at the large $l$ limit.

\section{Approximation of the dominant near-extremal modes}\label{appC}
In \cite{Hod:2017gvn} it has been proven that long-lived modes (or quasi-bound states) can be supported by a 4-dimensional near-extremal RN BH. In \cite{Cardoso1} it was realized that this family of modes exist in near-extremal RNdS BHs and is weakly dependent on the choice of $\Lambda$. Particularly, for neutral massless scalar fields, these modes can be very well approximated as
\begin{equation}
\omega_\text{d=4,NE}=-i(l+n+1)\kappa_-=-i(l+n+1)\kappa_+
\end{equation}
when $r_-=r_+$. Motivated by this result, we realize that for any dimension the dominant NE modes should be approximated by Eq. (\ref{NE}). For the sake of proving the validity of our approximate results, in Table \ref{table3} we show various dominant NE modes extracted from our spectral code versus the approximate Eq. (\ref{NE}). Higher overtones and angular numbers are not approximated by (\ref{NE}) anymore, but in any case, they are subdominant, thus they do not play any role for SCC.

\begin{table}[!htbp]
\centering
\begin{tabular}{||c|c|c|c||}
\hline
$\beta_\text{NE}$& $d=4$& $d=5$& $d=6$\\
\hline
$\beta_\text{approx}$&1 & 1 & 1\\
\hline
$\beta_\text{spectral}\,\,(l=n=0)$& 0.996&0.997&0.999\\
\hline
\end{tabular}
\caption{Comparison of $\beta_\text{approx}$ derived from (\ref{NE}) versus $\beta_\text{spectral}$ obtained with a spectral method for a $d$-dimensional RNdS BH with $M=1$, $\Lambda=0.1$ and $Q/Q_\text{max}=0.999999$.}\label{table3}
\end{table}
\end{appendix}
\bibliography{qnm}
\end{document}